\documentclass[12pt]{article}
\usepackage[centertags]{amsmath}
\usepackage{amsbsy}
\usepackage{amssymb}
\usepackage{epsfig}
\newcounter{figs}
\begin{document}
\begin{titlepage}
\null
\vspace{5mm}
\begin{flushright}
\begin{tabular}{l}
DFTT 3/97\\
hep-ph/9701298\\
January 1997
\end{tabular}
\end{flushright}
\vspace{3mm}
\begin{center}
\Large
\textbf{CONSTRAINTS ON NEUTRINOLESS DOUBLE BETA DECAY
FROM NEUTRINO OSCILLATION EXPERIMENTS}\\[3mm]
\normalsize
S.M. Bilenky\\
Joint Institute for Nuclear Research, Dubna, Russia, and\\
Technion, Physics Department, 32000 Haifa, Israel,\\[3mm]
C. Giunti\\
INFN, Sezione di Torino, and Dipartimento di Fisica Teorica,
Universit\`a di Torino,\\
Via P. Giuria 1, I--10125 Torino, Italy,\\[3mm]
and\\[3mm]
M. Monteno\\
INFN, Sezione di Torino, and Dipartimento di Fisica Sperimentale,
Universit\`a di Torino,\\
Via P. Giuria 1, I--10125 Torino, Italy.\\[3mm]
\textbf{Abstract}\\[3mm]
\begin{minipage}{0.8\textwidth}
We show that,
in the framework of
a general model with
mixing of three Majorana neutrinos
and a neutrino mass hierarchy,
the results of the Bugey and Krasnoyarsk
reactor neutrino oscillation experiments
imply strong limitations for the effective Majorana mass
$ | \left\langle m \right\rangle | $
that characterizes the amplitude of
neutrinoless double beta decay. 
We obtain further
limitations on
$ | \left\langle m \right\rangle | $
from
the data of the atmospheric neutrino experiments.
We discuss
the possible implications
of the results of the future long baseline
neutrino oscillation experiments
for neutrinoless double-$\beta$ decay.
\end{minipage}
\end{center}
\end{titlepage}

\section{Introduction}

The investigation of the fundamental properties of neutrinos
(neutrino masses, neutrino mixing,
the nature of neutrinos (Dirac or Majorana?)),
is the most important problem of today's neutrino physics.
This investigation is one of the major directions
of search for 
physics beyond the Standard Model.

At present there are several experimental 
indications in favor of neutrino oscillations.
The first indication was found in 
solar neutrino experiments
(Homestake,
Kamiokande,
GALLEX
and SAGE
\cite{solarexp}).
The second indication
was found in
the
Kamiokande
\cite{Kamiokande-atmospheric},
IMB and Soudan
\cite{IMB-Soudan}
atmospheric neutrino experiments.
A third indication in favor of neutrino oscillations
was claimed recently by the LSND
collaboration
\cite{LSND}.
On the other hand,
in many experiments with 
neutrinos
from reactors and accelerators
no indication in favor of neutrino
oscillations was found
(see the reviews in Refs.\cite{Boehm-Vannucci}).

The neutrino oscillation experiments
do not allow to answer to the question:
what type of particles are massive neutrinos,
Dirac or Majorana? 
(see Ref.\cite{BHP80}).
The answer to this question,
that
has a fundamental importance for the theory,
could be obtained
in experiments
on the search for processes in which
the total lepton number 
$ L=L_{e}+L_{\mu}+L_{\tau} $
is not conserved.
The classical process of this type is
neutrinoless double-$\beta$ decay
($(\beta\beta)_{0\nu}$):
\begin{equation}
(A,Z)
\to
(A,Z+2)
+
e^{-}
+
e^{-}
\;,
\label{01}
\end{equation}
The observation of this process would be an unambiguous
proof that neutrinos are massive Majorana particles.

At present,
the neutrinoless double-$\beta$ decay
of several nuclei
is searched for in more than 40 experiments
(see, for example, Ref.\cite{BBreviews}).
No positive indication in favor of this process was found
up to now.
The most stringent limits on the half-lives
for $(\beta\beta)_{0\nu}$ decay were found in
$^{76}$Ge and $^{136}$Xe experiments.
In the $^{76}$Ge experiments
of the Heidelberg-Moscow
and
IGEX
collaborations
\cite{Heidelberg-Moscow-IGEX}
it was found that
\begin{align}
\null & \null
T_{1/2}(^{76}\mathrm{Ge})
>
7.4 \times 10^{24} \, \mathrm{y}
\qquad \mbox{(90\% CL)}
\qquad
\mbox{Heidelberg-Moscow}
\;,
\label{021}
\\
\null & \null
T_{1/2}(^{76}\mathrm{Ge})
>
4.2 \times 10^{24} \, \mathrm{y}
\qquad \mbox{(90\% CL)}
\qquad
\mbox{IGEX}
\;.
\label{022}
\end{align}
In the $^{136}$Xe experiment of the Caltech-Neuchatel-PSI
collaboration
\cite{Caltech-Neuchatel-PSI}
it was found that
\begin{equation}
T_{1/2}(^{136}\mathrm{Xe})
>
4.2 \times 10^{23} \, \mathrm{y}
\qquad \mbox{(90\% CL)}
\;.
\label{03}
\end{equation}

There are different mechanisms of violation of the lepton number
that can be responsible for
$(\beta\beta)_{0\nu}$ decay
(see, for example, Ref.\cite{Mohapatra95}).
We will consider here
the contribution to the amplitude of the
$(\beta\beta)_{0\nu}$ process
due to 
the usual mechanism
of Majorana neutrino mixing.
This mechanism is based on the assumption that
the left-handed
flavor neutrino fields
$\nu_{{\ell}L}$
that enter in the standard
CC weak interaction Hamiltonian
\begin{equation}
\mathcal{H}_{I}
=
\frac{ G_{F} }{ \sqrt{2} }
\,
2
\sum_{\ell=e,\mu,\tau}
\overline{\ell}_{L}
\,
\gamma^{\alpha}
\,
\nu_{{\ell}L}
\, \,
j_{\alpha}^{\mathrm{CC}}
+
\mathrm{h.c.}
\label{04}
\end{equation}
(here $G_{F}$ is the Fermi constant,
$j_{\alpha}^{\mathrm{CC}}$ is the standard CC hadronic current)
are superpositions of
the left-handed components
$\nu_{iL}$
of massive Majorana neutrino fields:
\begin{equation}
\nu_{{\ell}L}
=
\sum_{i}
U_{{\ell}i}
\,
\nu_{iL}
\qquad \qquad
(\ell=e,\mu,\tau)
\;.
\label{05}
\end{equation}
Here 
$ \nu_{i} = \nu_{i}^{c} \equiv \mathcal{C} \overline{\nu}_{i}^{T} $
is the field of a Majorana neutrino with
mass $m_{i}$
($\mathcal{C}$ is the charge conjugation matrix)
and $U$ is the unitary mixing matrix.

In the framework of Majorana neutrino mixing,
$(\beta\beta)_{0\nu}$ decay
is a process of the second order in
the weak interaction,
with a virtual neutrino.
In the case of small neutrino
masses ($\lesssim$ 1 MeV),
the contribution
to the matrix element of
$(\beta\beta)_{0\nu}$ decay
of the left-handed weak interaction
is proportional to
the effective mass
(see, for example, Ref.\cite{BP87-CWKim93})
\begin{equation}
\left\langle m \right\rangle
=
\sum_{i}
U_{ei}^2
\,
m_{i}
\;.
\label{07}
\end{equation}

The negative results of the experiments
on the search for
$(\beta\beta)_{0\nu}$ decay
imply upper bounds for the
the parameter
$ | \left\langle m \right\rangle | $.
The numerical values of the upper bounds
depend on the model that is used for the calculations
of the nuclear matrix elements.
From
the results of the
$^{76}$Ge
and
$^{136}$Xe
experiments 
the following upper bounds were obtained:
\begin{alignat}{5}
\null & \null
| \left\langle m \right\rangle |
<
( 0.6 - 1.6 ) \, \mathrm{eV}
\null & \null
\qquad
\null & \null
\qquad
\mbox{($^{76}$Ge \cite{Heidelberg-Moscow-IGEX,Faessler96})}
\;,
\label{081}
\\
\null & \null
| \left\langle m \right\rangle |
<
( 2.3 - 2.7 ) \, \mathrm{eV}
\null & \null
\qquad
\null & \null
\qquad
\mbox{($^{136}$Xe \cite{Caltech-Neuchatel-PSI})}
\;.
\label{09}
\end{alignat}
A large progress of the experiments
on the search for
neutrinoless double-$\beta$ decay
is expected in the future.
Several collaborations plan to reach
a sensitivity of
$ 0.1 - 0.3 \, \mathrm{eV} $
for
$ | \left\langle m \right\rangle | $
\cite{Heidelberg-Moscow-IGEX,futureBB}.

In the present paper
we will obtain limits on the effective Majorana mass
$ | \left\langle m \right\rangle | $
from the existing results of neutrino oscillation experiments
under the assumption that
neutrinos with definite masses are Majorana particles.
Implications of the results of solar and reactor
neutrino experiments for
neutrinoless double-$\beta$ decay
were discussed
in Refs.\cite{PS94,BBGK}.
Here we will extend these considerations
and we will enlarge considerably
the range of possible values of the heaviest neutrino
mass.
We will show that
rather strong limitations on the parameter
$ | \left\langle m \right\rangle | $
can be obtained
from the results of
the reactor neutrino experiments.
We will also take into account the atmospheric neutrino anomaly
and
we will discuss the possible implications for
$(\beta\beta)_{0\nu}$ decay
of the results of the new data-taking
long baseline reactor neutrino experiments
CHOOZ and Palo Verde
\cite{CHOOZ-PaloVerde}.

\section{Mixing of three neutrinos with a mass hierarchy}

The results of the LEP experiments
on the measurement of the invisible 
width of the $Z$ boson
imply that only three flavor neutrinos exist in nature
(see Ref.\cite{RPP}).
The number of massive Majorana neutrinos 
that corresponds to three neutrino flavors
is equal to three in the case of 
a left-handed
Majorana mass term and can be more than three
in the general
case of a Dirac and Majorana mass term
(see, for example, Ref.\cite{BP87-CWKim93}).

We will consider the  case of three light Majorana
neutrinos. 
As it is well known,
a characteristic
feature of the mass spectra of
leptons, up and down quarks
is the hierarchy of the masses of the particles
of different generations.
What about neutrinos?
Different possibilities for
the mass spectrum
of three neutrinos were considered recently in the
literature
\cite{spectrum,BGKP}.
We will assume here that
the neutrino masses
$m_1$, $m_2$, $m_3$,
like the masses of quarks and leptons,
satisfy a hierarchy relation:
\begin{equation}
m_1 \ll m_2 \ll m_3
\;.
\label{10}
\end{equation}
This scheme corresponds to the standard
see-saw mechanism of neutrino mass generation,
which
is based on the assumption
that the total lepton number
is violated at a very large energy scale
and is the only known mechanism that
explains naturally the smallness of the neutrino masses
with respect to the lepton masses.
We will not 
assume,
however,
any specific see-saw relation between neutrino masses.
We will use only the results
of the neutrino oscillation experiments
in the general framework of a hierarchy 
of neutrino masses.

In all four solar neutrino experiments
(Homestake,
Kamiokande,
GALLEX
and SAGE
\cite{solarexp})
the detected event rates are significantly
smaller than the event rates predicted by the
Standard Solar Model (SSM)
\cite{SSM}.
Moreover,
a phenomenological analysis
of the data of the different solar neutrino experiments,
in which the predictions of the SSM are not used,
strongly suggest that
the solar neutrino problem is real
\cite{phenomenological}.
In order to take into account
the results of solar
neutrino experiments
in the framework of a
hierarchy of neutrino masses,
we have to assume 
that
$
\Delta{m}^2_{21}
\equiv
m_2^2 - m_1^2
$
is relevant for the suppression
of the flux of solar $\nu_e$'s.

If the disappearance of solar $\nu_e$'s
is due to neutrino mixing, 
the results of the solar neutrino experiments
and the predictions of the SSM
can be reconciled with
\begin{equation}
\Delta{m}^{2}_{21}
\sim
( 0.3 - 1.2 ) \times 10^{-5}\, \mathrm{eV}^2
\qquad
\mbox{or}
\qquad
\Delta{m}^{2}_{21} \sim 10^{-10}\, \mathrm{eV}^2
\;,
\label{11}
\end{equation}
in the case of
MSW resonant transitions
\cite{SOLMSW}
and
just-so vacuum oscillations
\cite{SOLVAC},
respectively.

Under the assumption of a neutrino mass hierarchy,
for the effective Majorana mass
$ | \left\langle m \right\rangle | $
that characterizes the amplitude of
$(\beta\beta)_{0\nu}$ decay
we have
\begin{equation}
| \left\langle m \right\rangle |
\simeq
| U_{e3} |^2 \, m_3
\simeq
| U_{e3} |^2 \, \sqrt{ \Delta{m}^{2} }
\;,
\label{13}
\end{equation}
with
$
\Delta{m}^{2}
\equiv
m_3^2 - m_1^2
$.

\section{Reactor and solar neutrinos}

In order to obtain
information on
the effective Majorana mass
$ | \left\langle m \right\rangle | $
from the results of
neutrino oscillation experiments,
we will use the method developed in
Ref.\cite{BBGK}
(see also Ref.\cite{PS94}).

In the case of a small 
$ \Delta{m}^{2}_{21} $
and a neutrino mass hierarchy, 
the modulus of the amplitude
$ \mathcal{A}_{\nu_{\ell}\to\nu_{\ell'}} $
of
the transitions
$ \nu_{\ell} \to \nu_{\ell'} $
of terrestrial neutrinos
is given by
\begin{equation}
\left|
\mathcal{A}_{\nu_{\ell}\to\nu_{\ell'}}
\right|
=
\left|
\delta_{{\ell'}{\ell}}
+
U_{{\ell'}3}
\,
U_{{\ell}3}^{*}
\left(
\mathrm{e}^{ \textstyle - i \frac{ \Delta{m}^{2} L }{ 2 p } }
-
1
\right)
\right|
\;.
\label{14}
\end{equation}
Here $L$ is the distance between
the neutrino source and the detector
and
$p$ is the neutrino momentum.
In Eq.(\ref{14}) we took into account the fact
that for the distances and energies
of terrestrial neutrino oscillation experiments
we have
\begin{equation}
\frac{ \Delta{m}^{2}_{21} L }{ 2 p }
\ll
1
\;.
\label{15}
\end{equation}
For the $\nu_{\ell}$
($\bar\nu_{\ell}$)
survival probability,
from Eq.(\ref{14}) we have
(see Ref.\cite{BBGK})
\begin{equation}
P_{\nu_{\ell}\to\nu_{\ell}}
=
P_{\bar\nu_{\ell}\to\bar\nu_{\ell}}
=
1
-
\frac{ 1 }{ 2 }
\,
B_{{\ell};{\ell}}
\left(
1
-
\cos
\frac{ \Delta{m}^{2} L }{ 2 p }
\right)
\;,
\label{17}
\end{equation}
where the oscillation amplitudes
$B_{{\ell};{\ell}}$
are given by
\begin{equation}
B_{{\ell};{\ell}}
=
4
\left| U_{{\ell}3} \right|^2
\left(
1
-
\left| U_{{\ell}3} \right|^2
\right)
\;,
\label{19}
\end{equation}
with
$
|U_{e3}|^2
+
|U_{\mu3}|^2
+
|U_{\tau3}|^2
=
1
$,
because of the unitarity of the mixing matrix.

Several oscillation experiments with 
reactor $\overline{\nu}_e$'s
have been performed
in the latest years
(see the review in Ref.\cite{Boehm-Vannucci}).
No indication in favor of neutrino oscillations
were found in these experiments.
In our analysis
we will use the data of the recent
Krasnoyarsk and Bugey
\cite{Krasnoyarsk-Bugey}
experiments,
which give the most stringent
limits for the neutrino oscillation parameters.

We will consider
the square of the largest neutrino mass
$ m_3^2 \simeq \Delta{m}^{2} $
as an unknown 
parameter.
Taking into account
the limits for
the neutrino mass
obtained by the $^3$H $\beta$-decay experiments
(see Ref.\cite{RPP}),
we will consider
the region
$
\Delta{m}^{2}
\leq
10^{2} \, \mathrm{eV}^2
$.
The negative results
of the reactor neutrino oscillation experiments
allow to obtain an upper bound
for the effective Majorana mass
$ | \left\langle m \right\rangle | $
in the wide interval
\begin{equation}
10^{-2} \, \mathrm{eV}^2
\leq
\Delta{m}^{2}
\leq
10^{2} \, \mathrm{eV}^2
\;.
\label{20}
\end{equation}
At any fixed value of
$ \Delta{m}^{2} $
we have
an upper limit for the amplitude
$B_{{e};{e}}$
of
$ \bar\nu_{e} \to \bar\nu_{e} $
transitions
\begin{equation}
B_{{e};{e}}
\leq
B_{{e};{e}}^{0}
\;.
\label{21}
\end{equation}
The quantity
$ B_{{e};{e}}^{0} $
can be obtained
from the exclusion plots found from the data of reactor experiments.
From Eqs.(\ref{19}) and (\ref{21})
it follows that
$ \left| U_{e3} \right|^2 $
must satisfy one of the two inequalities:
\begin{align}
\null & \null
\left| U_{e3} \right|^2
\leq
a_{e}^{0}
\;,
\label{22}
\\
\mbox{or}
\quad
\null & \null
\nonumber
\\
\null & \null
\left| U_{e3} \right|^2
\geq
1 - a_{e}^{0}
\;,
\label{23}
\end{align}
where
\begin{equation}
a_{e}^{0}
\equiv
\frac{ 1 }{ 2 }
\left(
1
-
\sqrt{
1
-
B_{{e};{e}}^{0}
}
\right)
\;.
\label{24}
\end{equation}

In Fig.\ref{fig1}
we have plotted the values of
the parameter
$a_{e}^{0}$
obtained from
the 90\% CL
exclusion plots of the Bugey and Krasnoyarsk experiments,
for
$ \Delta{m}^{2} $
in  the range (\ref{20}).
Figure \ref{fig1}
shows that
$a_{e}^{0}$
is small
in the considered range of
$ \Delta{m}^{2} $.
Thus,
from the results of the reactor oscillation experiments 
it follows that
$\left| U_{e3} \right|^2$
can only be small
or large (close to one).

The results of the
solar neutrino experiments exclude this last possibility.
In fact,
the averaged probability
$ P_{\nu_e\to\nu_e}(E) $
of solar $\nu_e$'s
to survive,
in the case of a neutrino mass hierarchy with
$ \Delta{m}^{2}_{21} $
relevant for the oscillations of solar neutrinos,
is given by
(see Ref.\cite{SS92})
\begin{equation}
P_{\nu_e\to\nu_e}(E)
=
\left(
1
-
\left| U_{e3} \right|^2
\right)^2
P_{\nu_e\to\nu_e}^{(1,2)}(E)
+
\left| U_{e3} \right|^4
\;,
\label{25}
\end{equation}
where
$ P_{\nu_e\to\nu_e}^{(1,2)}(E) $
is the $\nu_{e}$ survival probability 
due to the mixing between
the first and the second generations
and
$E$ is the neutrino energy.
If
$\left| U_{e3} \right|^2$
satisfies the inequality (\ref{23}),
we have
$
P_{\nu_e\to\nu_e}(E)
\geq
( 1 - a_{e}^{0} )^2
\equiv
P_{\nu_e\to\nu_e}^{\mathrm{min}}
$.
In
Fig.\ref{fig2}
we have plotted the values of
$ P_{\nu_e\to\nu_e}^{\mathrm{min}} $
obtained from the
results of the Bugey and Krasnoyarsk experiments,
for
$ \Delta{m}^2 $
in the interval (\ref{20}).
It can be seen that
$ P_{\nu_e\to\nu_e}(E) > 0.65 $
for all the considered values of $ \Delta{m}^2 $
and
$ P_{\nu_e\to\nu_e}(E) > 0.91 $
for
$
3 \times 10^{-2} \, \mathrm{eV}^2
\leq
\Delta{m}^{2}
\leq
10^{2} \, \mathrm{eV}^2
$.
Furthermore,
Eq.(\ref{25}) implies that
the maximal variation of
$P_{\nu_e\to\nu_e}(E)$
as a function of neutrino energy
is given by
$ ( 1 - |U_{e3}|^2 )^2 $.
If
$ |U_{e3}|^2$
satisfies the inequality (\ref{23}),
we have
$ ( 1 - |U_{e3}|^2 )^2 \leq (a_{e}^{0})^2 $,
which is a very small quantity
(from Fig.\ref{fig1}
one can see that
$ ( 1 - |U_{e3}|^2 )^2 \leq 4 \times 10^{-2} $
for
$ \Delta{m}^2 $
in the interval (\ref{20})).
Thus,
in this case
$ P_{\nu_e\to\nu_e} $
as a function of neutrino energy
is practically constant.
The
large lower bound for
the survival probability
$ P_{\nu_e\to\nu_e} $
and its practical independence from the neutrino energy
are not compatible with the data of the solar neutrino experiments
(see Ref.\cite{KP96}).
Thus,
from the results of the solar 
and reactor neutrino experiments
we come to the conclusion
that
$\left| U_{e3} \right|^2$
is small and satisfies the
inequality (\ref{22}).

The limit (\ref{22}) for
$\left| U_{e3} \right|^2$
implies
the following upper bound for the effective Majorana mass
$ | \left\langle m \right\rangle | $:
\begin{equation}
| \left\langle m \right\rangle |
\leq
a_{e}^{0}
\,
\sqrt{ \Delta{m}^{2} }
\;.
\label{26}
\end{equation}
The upper bounds
obtained with Eq.(\ref{26}) from the
90\% CL exclusion plots
of the
Bugey and Krasnoyarsk
reactor neutrino oscillation experiments
are  presented in Fig.\ref{fig3}.
With the thick solid line 
we have drawn the unitarity upper bound
$
| \left\langle m \right\rangle |
\leq
\sqrt{ \Delta{m}^{2} }
$.

As it is seen from Fig.\ref{fig3},
from the results of
the reactor neutrino experiments it follows that
for
$ \Delta{m}^{2} \lesssim 10 \, \mathrm{eV}^2 $
the effective Majorana mass
$ | \left\langle m \right\rangle | $
cannot be larger than
$ 10^{-1} \, \mathrm{eV} $.
Let us stress that
the sensitivity to
$ | \left\langle m \right\rangle | \simeq 10^{-1} \, \mathrm{eV} $
is the goal of future experiments
on the search for
$(\beta\beta)_{0\nu}$ decay
\cite{Heidelberg-Moscow-IGEX,futureBB}.

In the region
$
10 \, \mathrm{eV}^2
\lesssim \Delta{m}^{2} \lesssim
10^2 \, \mathrm{eV}^2
$
the upper bound for the parameter
$ | \left\langle m \right\rangle | $
grows with
$ \Delta{m}^{2} $
and reaches the value of
$ 4 \times 10^{-1} \, \mathrm{eV} $
at
$ \Delta{m}^{2} \simeq 10^2 \, \mathrm{eV}^2 $.
The region
of relatively large values of
$ \Delta{m}^{2} $
($ 10 - 10^2 \, \mathrm{eV}^2 $)
is very important for 
the dark matter problem.
Two experiments at CERN
(CHORUS and NOMAD \cite{CHORUS-NOMAD})
are searching for
$ \nu_\mu \to \nu_\tau $
oscillations
in this range of 
$ \Delta{m}^{2} $.

In Fig.\ref{fig3} we have also presented the upper
bound for
$ | \left\langle m \right\rangle | $
that corresponds to the projected sensitivity
of the reactor long baseline neutrino oscillation experiments
CHOOZ and Palo Verde
\cite{CHOOZ-PaloVerde}.
These experiments
will allow to obtain new limits
for the effective Majorana mass
$ | \left\langle m \right\rangle | $
in the region of small $ \Delta{m}^{2} $.
If their sensitivity limit will be reached
without detecting neutrino oscillations,
in the region
$ \Delta{m}^{2} \lesssim 2 \, \mathrm{eV}^2 $
the upper bound for
$ | \left\langle m \right\rangle | $
will be less than about
$ 3 \times 10^{-2} \, \mathrm{eV} $.

\section{Atmospheric neutrinos}

Up to now we have taken into account only
the results of
solar and reactor neutrino experiments
and
we considered the heaviest neutrino mass
$ m_{3} \simeq \sqrt{ \Delta{m}^{2} } $
as an unknown parameter.
Let us take now into account also the results of the
atmospheric neutrino experiments.

The Kamiokande collaboration found
\cite{Kamiokande-atmospheric}
that the detected ratio of
muon and electron atmospheric neutrino events
is significantly smaller than the expected one.
For
the double ratio
$
R = (\mu/e)_{\mathrm{data}}/(\mu/e)_{\mathrm{MC}}
$
($(\mu/e)_{\mathrm{MC}}$
is the Monte-Carlo
calculated ratio of muon and electron events
under the assumption that neutrinos do not oscillate),
the Kamiokande collaboration found
\begin{equation}
R_{\mathrm{Kamiokande}}^{\mathrm{sub-GeV}}
=
0.60 \mbox{}^{+0.06}_{-0.05} \pm 0.05
\;,
\qquad
\qquad
R_{\mathrm{Kamiokande}}^{\mathrm{multi-GeV}}
=
0.57 \mbox{}^{+0.08}_{-0.07} \pm 0.07
\;.
\label{27}
\end{equation}
The atmospheric neutrino anomaly
was observed also in the
IMB and Soudan
\cite{IMB-Soudan}
experiments:
\begin{equation}
R_{\mathrm{IMB}}
=
0.54 \pm 0.05 \pm 0.12
\;,
\qquad
\qquad
R_{\mathrm{Soudan}}
=
0.75 \pm 0.16 \pm 0.10
\;.
\label{29}
\end{equation}
On the other hand,
the double ratio R measured
in the Frejus and NUSEX
\cite{Frejus-NUSEX}
experiments
is compatible with one:
\begin{equation}
R_{\mathrm{Frejus}}
=
0.99 \pm 0.13 \pm 0.08
\;,
\qquad
\qquad
R_{\mathrm{NUSEX}}
=
1.04 \pm 0.25
\;.
\label{31}
\end{equation}

The results of the
Kamiokande experiment can be explained
\cite{Kamiokande-atmospheric}
by two flavor neutrino oscillations
$ \nu_{\mu} \leftrightarrows \nu_{\tau} $
or
$ \nu_{\mu} \leftrightarrows \nu_{e} $.
with the following allowed ranges
for the oscillation parameters:
\begin{alignat}{5}
&
5 \times 10^{-3}
\lesssim
\Delta m^2
\lesssim
3 \times 10^{-2} \, \mbox{eV}^2
& \qquad &
0.7
\lesssim
\sin^2 2\vartheta
\lesssim
1
&
\qquad
(\nu_{\mu} \leftrightarrows \nu_{\tau})
\;,
\label{32}
\\
&
7 \times 10^{-3}
\lesssim
\Delta m^2
\lesssim
8 \times 10^{-2} \, \mbox{eV}^2
& \qquad &
0.6
\lesssim
\sin^2 2\vartheta
\lesssim
1
&
\qquad
(\nu_{\mu} \leftrightarrows \nu_{e})
\;.
\label{33}
\end{alignat}

We have analyzed
the Kamiokande atmospheric neutrino data
\cite{Kamiokande-atmospheric}
in the framework of the model under consideration
with mixing of three neutrinos and a neutrino mass hierarchy.
The oscillation probabilities of atmospheric neutrinos
depend on three parameters:
$\Delta{m}^{2}$,
$|U_{e3}|^2$
and
$|U_{\mu3}|^2$.
The matter effect
for the atmospheric neutrinos
reaching the Kamiokande detector from below
has been taken into account.
The presence of matter
is important because it modifies the phases
of neutrino oscillations
\cite{Pantaleone94}
and its effect is to enlarge
the allowed region towards low values of
$ \Delta{m}^{2} $.
The best fit of the Kamiokande data
is obtained for
\begin{equation}
\Delta{m}^{2} = 2.3 \times 10^{-2} \, \mathrm{eV}^2
\;,
\qquad
|U_{e3}|^2 = 0.26
\;,
\qquad
|U_{\mu3}|^2 = 0.49
\;,
\label{34}
\end{equation}
with
$ \chi^2 = 6.7 $
for 9 degrees of freedom,
corresponding to a CL of 67\%.
The range \emph{allowed} at 90\% CL
in the plane of the parameters
$ \Delta{m}^{2} $,
$ | \left\langle m \right\rangle | $
is shown in Fig.\ref{fig3}
as the region
enclosed by the dash-dotted curve.
The best fit of the Kamiokande data
corresponds to
$
| \left\langle m \right\rangle |
=
3.9 \times 10^{-2} \, \mathrm{eV}
$
(the triangle in Fig.\ref{fig3}).
From Fig.\ref{fig3}
it can be seen that
the results of
Kamiokande experiment imply that
\begin{equation}
| \left\langle m \right\rangle |
\lesssim
10^{-1} \, \mathrm{eV}
\;.
\label{35}
\end{equation}

If we take into account also the limit
obtained from the reactor neutrino oscillation experiments,
for upper bound
of the effective Majorana mass we have
\begin{equation}
| \left\langle m \right\rangle |
\lesssim
7 \times 10^{-2} \, \mathrm{eV}
\;.
\label{36}
\end{equation}

It is interesting to investigate how the region
in the plane of the parameters
$ \Delta{m}^{2} $,
$ | \left\langle m \right\rangle | $
allowed by the Kamiokande data
is modified by the inclusion in the fit
of the data obtained in the Frejus experiment.
The best fit
of Kamiokande and Frejus data
is obtained for
\begin{equation}
\Delta{m}^{2} = 1.7 \times 10^{-2} \, \mathrm{eV}^2
\;,
\quad
|U_{e3}|^2 = 0.17
\;,
\quad
|U_{\mu3}|^2 = 0.29
\;,
\label{37}
\end{equation}
with
$ \chi^2 = 28 $
for 19 degrees of freedom
corresponding to a CL of 8\%
(the best fit value of
$ | \left\langle m \right\rangle | $
is
$ 2.2 \times 10^{-2} \, \mathrm{eV} $,
depicted as a
square in Fig.\ref{fig3}).
The corresponding allowed region
(at 90\% CL)
is shown in Fig.\ref{fig3}
as the region enclosed
by the dash-dot-dotted curve.
The figure shows that
the region allowed by the combined
Kamiokande and Frejus data
is not very different from the region allowed
by the Kamiokande data alone.
If we take into account
the limit obtained from the results of the Bugey and Krasnoyarsk
reactor experiments,
we have
\begin{equation}
| \left\langle m \right\rangle |
\lesssim
4 \times 10^{-2} \, \mathrm{eV}
\;.
\label{38}
\end{equation}

The limits
on the effective Majorana mass
$ | \left\langle m \right\rangle | $
obtained above
could decrease substantially
after the fulfillment of the program 
of neutrino oscillation
experiments of the next generation,
that will explore the region
of neutrino mixing parameters allowed by the
data of the atmospheric neutrino experiments.
We are referring to the
Super-Kamiokande
\cite{SK}
and long baseline accelerator
(KEK--Super-Kamiokande
\cite{SK},
MINOS
and
ICARUS
\cite{MINOS-ICARUS})
and reactor
(CHOOZ and Palo Verde
\cite{CHOOZ-PaloVerde})
experiments.

If,
for example,
$ \nu_{\mu} \leftrightarrows \nu_{\tau} $
oscillations
with
$ \Delta{m}^2 \sim 10^{-2} \, \mathrm{eV}^2 $
will be found in the
accelerator long baseline experiments
and the reactor long baseline experiments
will reach their projected sensitivity,
the region of allowed
values of the parameters
$ | \left\langle m \right\rangle | $,
$ \Delta{m}^{2} $
will be very small
(see Fig.\ref{fig3}),
with
the upper bound
\begin{equation}
| \left\langle m \right\rangle |
\lesssim
10^{-2} \, \mathrm{eV}
\;.
\label{39}
\end{equation}
For the
$(\beta\beta)_{0\nu}$ decay
experiments
the sensitivity to
$
| \left\langle m \right\rangle |
\simeq
10^{-2} \, \mathrm{eV}
$
is a challenging problem
\cite{Snowmass94}.

On the other hand,
if the
atmospheric neutrino anomaly will be confirmed
in the
$ \nu_{\mu} \leftrightarrows \nu_{e} $
channel
by reactor long baseline neutrino oscillation experiments,
it will mean that the value of
$ | \left\langle m \right\rangle | $
lies within the
interval
$ 10^{-3} - 10^{-1} \, \mathrm{eV} $.
In this case,
from a determination of
$ | U_{e3} |^2 $
and
$ \Delta{m}^{2} $
in reactor experiments,
through Eq.(\ref{13})
it will be possible to infer
the value of
$ | \left\langle m \right\rangle | $.

\section{Conclusions}

We have shown that,
in the framework of
a general model with
mixing of three Majorana neutrinos
and a neutrino mass hierarchy,
from the results of the reactor,
solar and
atmospheric neutrino experiments
it is possible to obtain
rather strict limits
on the value of the effective Majorana mass
$ | \left\langle m \right\rangle | $,
that characterizes the amplitude of
neutrinoless double beta decay.

We have shown that the results of
the Bugey and Krasnoyarsk reactor neutrino oscillation experiments
imply that 
$
| \left\langle m \right\rangle |
\lesssim
10^{-1} \, \mathrm{eV}
$
if
$ \Delta{m}^{2} $
is less than about
$ 10 \, \mathrm{eV}^2 $.
If
$ \Delta{m}^{2} $
has a value in the interval
$ 10 - 100 \, \mathrm{eV}^2 $,
we have
$
| \left\langle m \right\rangle |
\lesssim
4 \times 10^{-1} \, \mathrm{eV}
$.

We have also shown that
the Kamiokande 
atmospheric neutrino data imply
that
$
| \left\langle m \right\rangle |
\lesssim
7 \times 10^{-2} \, \mathrm{eV}
$.
Future long baseline reactor and accelerator
neutrino experiments could decrease the upper bound
for the effective Majorana mass
$ | \left\langle m \right\rangle | $
to the level of about
$ 10^{-2} \, \mathrm{eV} $.

The constraints 
on the value of the effective Majorana mass
$ | \left\langle m \right\rangle | $
that follow from neutrino oscillation experiments
must be taken into account
in the interpretation of the data of
$(\beta\beta)_{0\nu}$ decay
experiments.
The observation in the future 
experiments of neutrinoless double-$\beta$ decay with 
a half-life that corresponds to a value
of the effective mass
$ | \left\langle m \right\rangle | $
that is
significantly larger than 
$ 10^{-1} \, \mathrm{eV} $
could imply that the mass $m_3$ of the heaviest
neutrino is larger than
$ 2 - 3 \, \mathrm{eV} $,
or that
the neutrino masses 
do not have a hierarchy pattern\footnote{
Another possibility for the spectrum
of three neutrinos that can accommodate
the data of solar neutrino experiments
is the inverted mass hierarchy
$ m_1 \ll m_2 \simeq m_3 $.
In this case the reactor data 
do not imply any limitations
on the effective Majorana mass
$ | \left\langle m \right\rangle | $
\cite{BGKP}.
}
and
are not generated by the standard see-saw mechanism.
Other possibilities are
that a new non-standard interaction 
is responsible for
neutrinoless double-$\beta$ decay
(right-handed currents, \ldots,
for a review see Ref.\cite{Mohapatra95}),
or that four or
more massive Majorana neutrinos exist in nature\footnote{
The schemes with four massive neutrinos allow to accommodate
all existing indications in favor of neutrino mixing
\cite{four,BGKP,BGG},
including the indications
obtained recently in
the LSND experiment \cite{LSND}.
The connection between
$(\beta\beta)_{0\nu}$ decay
and
neutrino oscillations
in the framework of schemes with four massive
neutrinos
was discussed
in Refs.\cite{PS94,BGKP,BGG}.
Let us notice that the LSND indications in favor of
$ \nu_{\mu} \leftrightarrows \nu_{e} $
oscillations will be tested 
in two years by
the KARMEN experiment
\cite{KARMEN}.
}.

\begin{flushleft}
\Large \textbf{Acknowledgments}
\end{flushleft}

This work was done while one of
authors (S.B) was  
Lady Davis visiting professor at the Technion.
The author would like to thank
the Physics Department of Technion 
for its hospitality and A. Dar and D. Wyler for
useful discussions.

\newpage

\begin{flushleft}
\Large \textbf{Figure Captions}
\end{flushleft}
\begin{description}

\item[Figure~\ref{fig1}.]
\refstepcounter{figs} \label{fig1}
The quantity $a_{e}^{0}$
(see Eq.(\ref{24}))
obtained from
the 90\% CL exclusion plots of the
Bugey and Krasnoyarsk
\cite{Krasnoyarsk-Bugey}
reactor neutrino oscillation experiments.

\item[Figure~\ref{fig2}.]
\refstepcounter{figs} \label{fig2}
The 90\% CL lower bound for the probability
of solar $\nu_e$'s to survive in the 
case of a large value of the parameter
$|U_{e3}|^2$
($ \geq 1 - a_{e}^{0} $).

\item[Figure~\ref{fig3}.]
\refstepcounter{figs} \label{fig3}
The 90\% CL upper bound for
the effective Majorana mass
$ | \left\langle m \right\rangle | $
obtained from the results of
the Bugey (solid line) and Krasnoyarsk (dotted line)
neutrino reactor experiments.
The regions
enclosed by dash-dotted ( dash-dot-dotted) lines
are allowed
by the data of
the Kamiokande (Kamiokande and Frejus) 
atmospheric neutrino experiments.
The upper bound for
$ | \left\langle m \right\rangle | $
that corresponds to
the projected sensitivity of the long baseline
CHOOZ (short-dashed line) and Palo Verde (long-dashed line)
reactor neutrino oscillation experiments are also shown.

\end{description}

\begin{figure}[p]
\begin{center}
\mbox{\epsfig{file=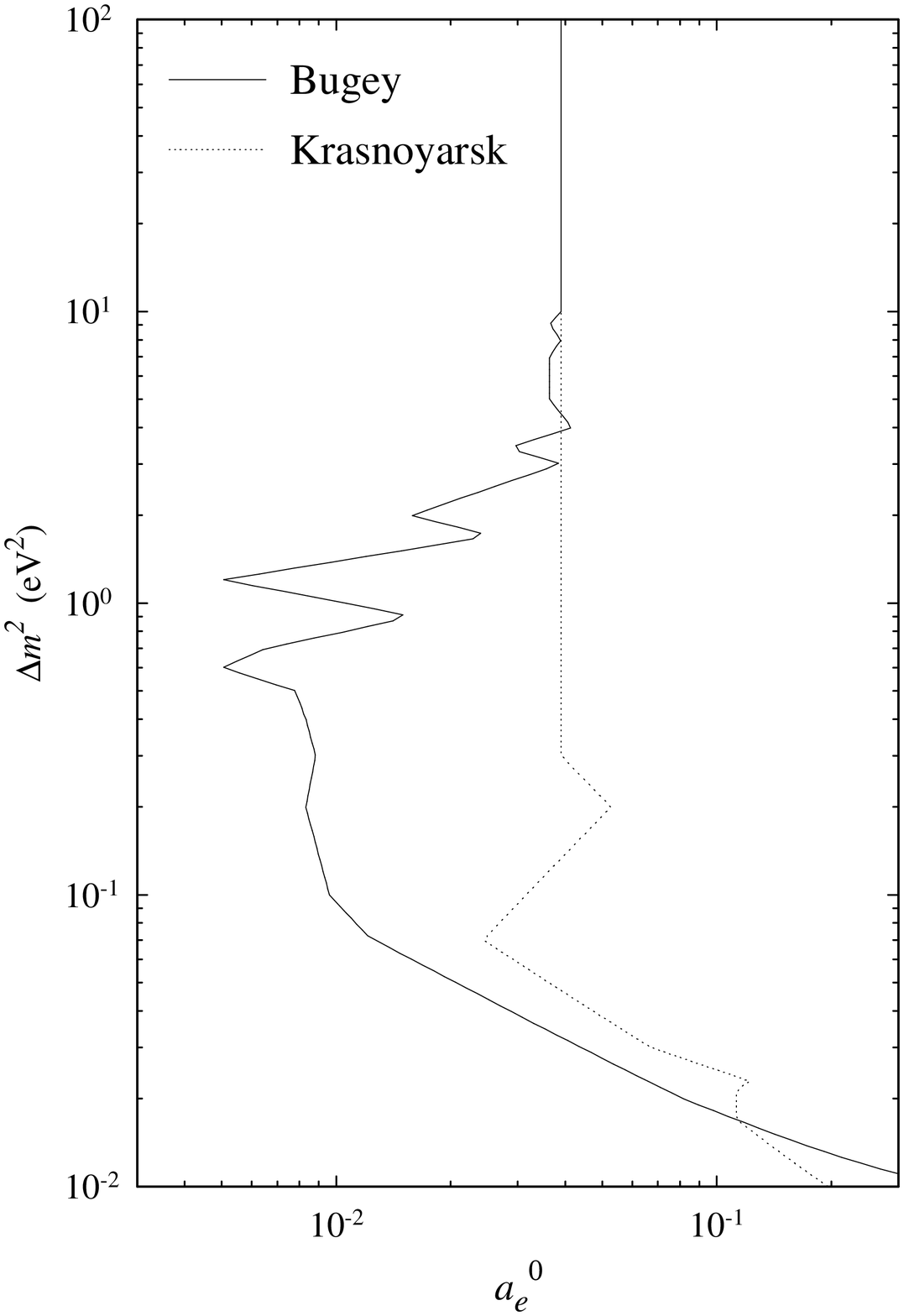,width=0.79\textwidth}}
\\
\Large Figure~\ref{fig1}
\end{center}
\end{figure}

\begin{figure}[p]
\begin{center}
\mbox{\epsfig{file=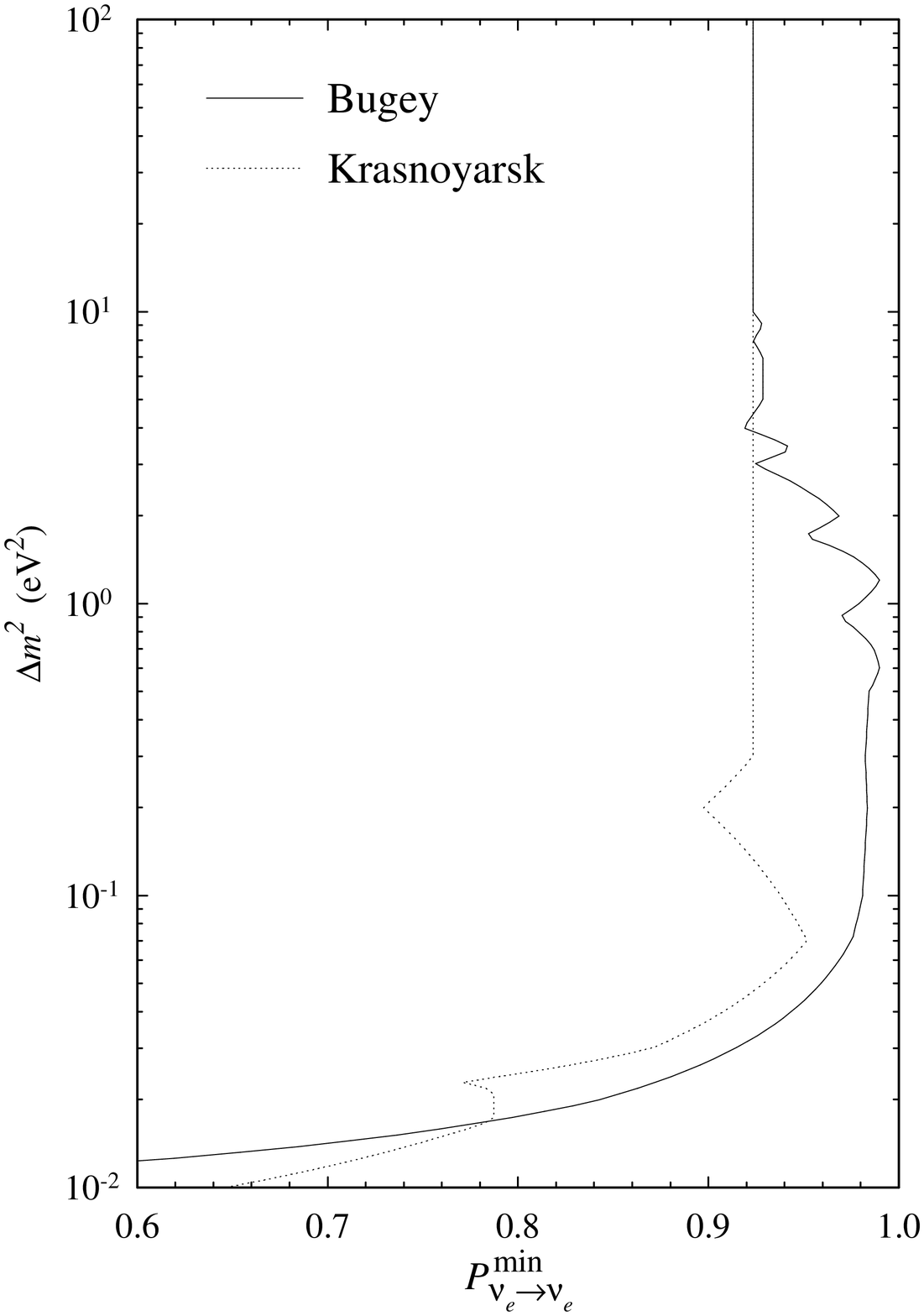,width=0.79\textwidth}}
\\
\Large Figure~\ref{fig2}
\end{center}
\end{figure}

\begin{figure}[p]
\begin{center}
\mbox{\epsfig{file=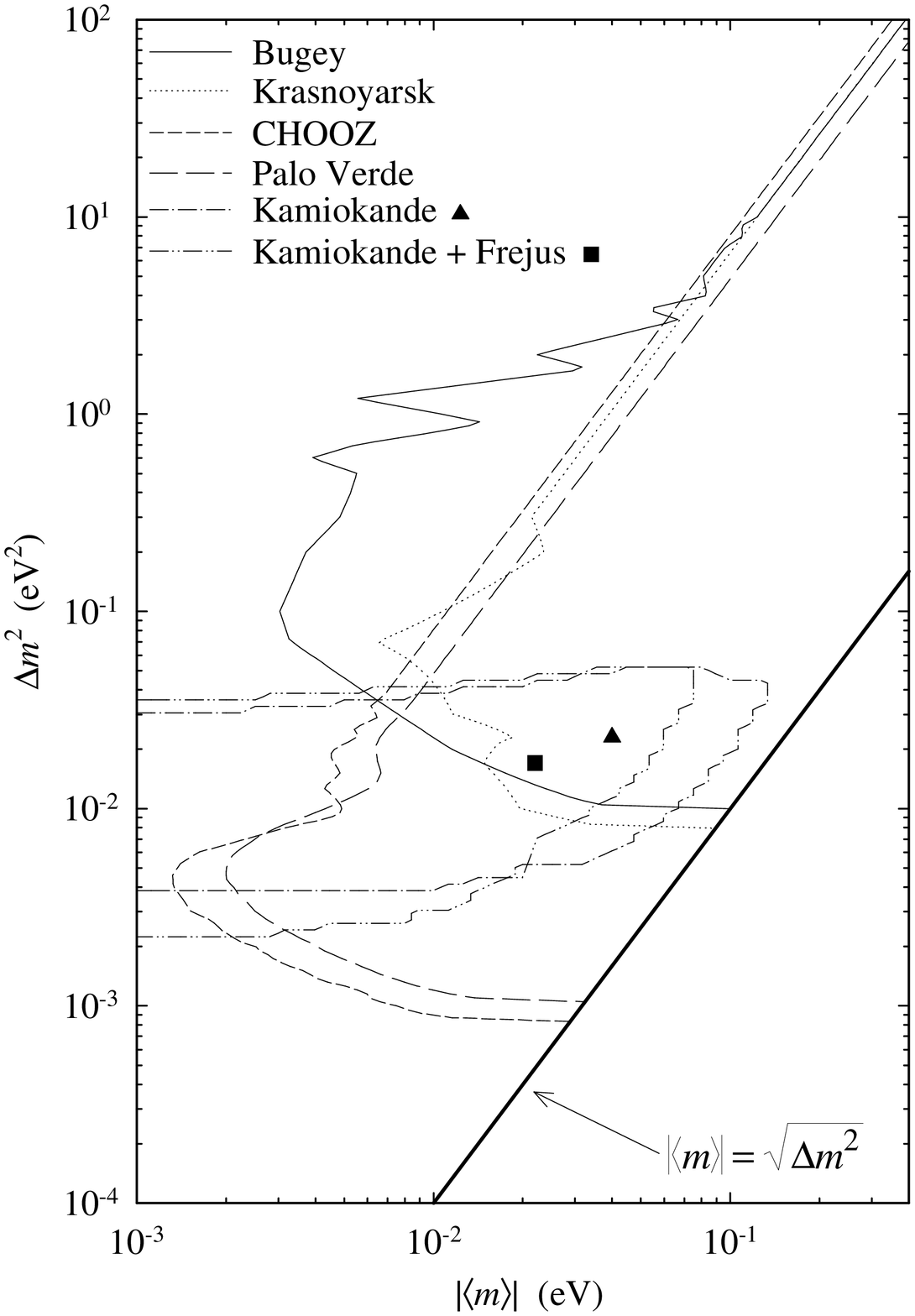,width=0.79\textwidth}}
\\
\Large Figure~\ref{fig3}
\end{center}
\end{figure}

\end{document}